# Search for a quantum phase transition in U(Pt$_{1-x}$ Pd$_x$)$_3$

M. J. Graf [a,1], R. J. Keizer [b], A. de Visser [b], S. T. Hannahs [c]

[a] Department of Physics, Boston College, Chestnut Hill, MA 02459, USA
[b] Van der Waals-Zeeman Inst., Univ. of Amsterdam, 1018XE Amsterdam, The Netherlands
[c] National High Magnetic Field Lab., Florida State University, Tallahassee, FL 32310, USA

**Abstract**

Pd in U(Pt$_{1-x}$ Pd$_x$)$_3$ suppresses the superconducting T$_c$ to 0 K at x$_c \simeq 0.007$ and induces a conventional AFM state for x $\geq$ x$_c$. The resistivity below 1 K for x $\leq 0.02$ shows a deviation from Fermi liquid behavior described by $\rho(T) = \rho_0 + AT^\alpha$; $\alpha$ varies from 2 for x = 0 to 1.6 for x $\simeq$ x$_c$. This suggests that a quantum phase transition (QPT) exists near x$_c$. Transport for a sample with x = 0.004 < x$_c$ has a pressure-independent exponent $\alpha = 1.77$, suggesting that if a QPT exists it may be associated with the magnetic transition.

*Keywords:* UPt$_3$; quantum phase transition; unconventional superconductivity; heavy fermions

Pd-substitution is a powerful technique for studying superconductivity, magnetism, and their interplay in UPt$_3$. It is the only known way to increase the splitting of the double superconducting transition.[1] This increase has been correlated with an increase in the ordered moment associated with anomalous small-moment antiferromagnetism (SMAF).[2] Also, it has been shown that Pd-substitution both suppresses the superconducting T$_c$ to 0 K at x$_c \simeq 0.007$[3] and induces conventional large-moment antiferromagnetism (LMAF) for x > x$_c$;[4] the phase diagram for x $\leq$ 0.02, from resistivity[3] and $\mu$SR measurements,[5] is shown in Fig. 1. It is strongly suggestive of a competition between superconductivity (SC) and static magnetic ordering (LMAF), as expected for a spin-fluctuation-based pairing mechanism.

[1] Corresponding author. E-mail: grafm@bc.edu

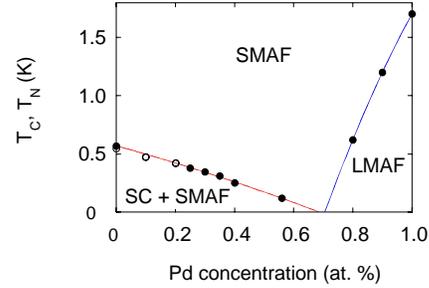

Fig. 1. U(Pt$_{1-x}$ Pd$_x$)$_3$ phase diagram, x $\leq$ 0.02 (open and solid symbols for single and polycrystals, respectively).

Apart from the nature of, and distinction between, the SMAF and LMAF phases, the phase diagram in Fig. 1 raises interesting questions. A quantum phase transition (QPT) may occur if the phase lines indeed go to T = 0 K near x = 0.007. The QPT could be of magnetic origin, as observed in other heavy fermion systems[6], or possibly

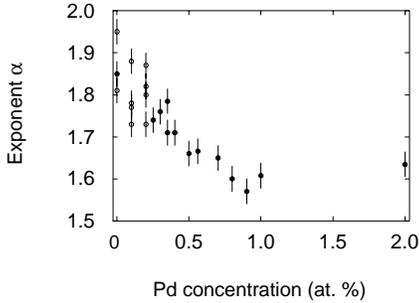

Fig. 2. Power law exponent versus Pd concentration.

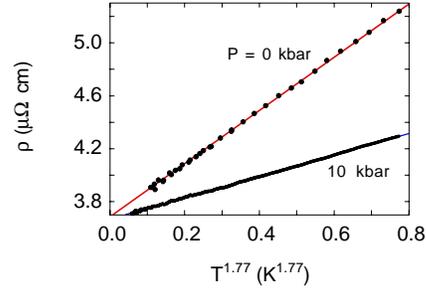

Fig. 3. Pressure-dependence of the resistivity, x = 0.004.

be associated with superconductivity.[7] Here we present initial studies of these possibilities.

First, we examine the temperature-dependent resistivity of U(Pt$_{1-x}$ Pd$_x$)$_3$ for x $\leq$ 0.02 and for T $\leq$ 1 K for a variety of polycrystal and single crystal samples. Pure UPt$_3$ has a Fermi liquid-like low-T resistivity with a quadratic T-dependence. As Pd is substituted in for Pt, we observe a clear deviation from quadratic behavior. The quadratic term is thought to arise from spin-fluctuation scattering, and the resistivity can be written $\rho(T) = \rho_0 + A(T/T_{sf})^2$, where $T_{sf}$ is the spin-fluctuation temperature (roughly 18 K in pure UPt$_3$[8]). This holds only when T $\ll T_{sf}$. The observed deviation could be explained within a Fermi liquid picture if $T_{sf}$ was reduced by well over a factor of two for Pd concentrations of x = 0.005; this is inconsistent with thermodynamic measurements.

The data is best described by $\rho(T) = \rho_0 + AT^\alpha$, with $\alpha$ varying from 2 for x = 0 to 1.6 for x $\simeq$ x$_c$; from limited data above x = 0.01 it appears that $\alpha$ either stays constant, or increases weakly, for x > x$_c$ (see Fig. 2). This suggests that a quantum phase transition (QPT) exists near x$_c$, associated with either T$_c$ or the Néel temperature T$_N$ approaching 0 K. The value 1.6 is near the predicted value of 1.5 for 3-D critical fluctuations with dynamic exponent z = 2.[9]

Transport data for a polycrystalline sample with x = 0.004 < x$_c$ is shown in Fig. 3 for ambient pressure (T$_c$ = 0.25 K) and 10 kbar. Data for the suppression of T$_c$ will be presented elsewhere, but T$_c$ approaches 0 K at 10 kbar. While the coefficient of the temperature dependent resistivity is reduced, the exponent $\alpha$ = 1.77 is pressure-independent, suggesting that if a QPT exists it may be associated with the magnetic transition. We are currently studying the pressure-dependent transport of samples with x $\geq$ x$_c$ to determine the change in $\alpha$ as the LMAF T$_N$ approaches 0 K.

## Acknowledgement

Work was supported by Research Corporation (RA0246), NATO (CG11096) and by the Dutch funding agency FOM; the NHMFL is sponsored by the NSF and the State of Florida.